\documentclass{moriond}

\def\be{\begin{equation}}
\def\ee{\end{equation}}
\def\bea{\begin{eqnarray}}
\def\eea{\end{eqnarray}}

\newcommand{\Photo}{}

\newcommand{\popII}{PopII}%
\newcommand{\popIII}{PopIII}%
\begin{document}
\vspace*{4cm}
\title{Future 21cm constraints on dark matter energy injection: Application to ALPs}

\author{ Laura Lopez Honorez}

\address{Service de Physique Th\'eorique, C.P. 225, Universit\'e Libre de Bruxelles,\\ Boulevard du Triomphe,  and Theoretische Natuurkunde \& The International Solvay Institutes, \\ Vrije Universiteit Brussel, Pleinlaan 2, B-1050 Brussels, Belgium}

\maketitle\abstracts{ The redshifted 21cm signal from the Cosmic Dawn is expected to provide unprecedented insights into early Universe astrophysics and cosmology. Here, we briefly summarize how decaying dark matter can heat the intergalactic medium before the first galaxies, leaving a distinctive imprint on the 21cm power spectrum. We discuss the first Fisher matrix forecasts on the sensitivity of the Hydrogen Epoch of Reionization Array telescope (HERA) and argue that HERA can improve by up to three orders of magnitude previous cosmology constraints. In these proceedings, we project these future bounds in the plane of the Axion like particles (ALP) coupling to photons as a function of the ALP mass. We focus on the ALP mass range between $\sim$ 10 keV and 1 MeV where the 21cm signal power spectrum probes are expected to improve on any other current dark matter searches. This illustrates how 21cm cosmology can be expected to help in probing uncharted regions of the dark matter parameter space beyond the reach of existing astro-particle and cosmology experiments.}

\section{Introduction}
\label{sec:introduction}

Dark matter (DM) is a crucial component of our Universe that shapes
its evolution. Cosmological probes are amongst the powerful tools at
our disposal to shed light on the nature of DM. In particular, the
Cosmic Microwave Background (CMB)~\cite{Planck:2018vyg} currently
provides the strongest constraints on DM properties probing its energy
density today to the percent level. The CMB can also constrain DM
annihilation or decay into SM
particles~\cite{Planck:2018vyg,Capozzi:2023xie}. Both scenarios result
in an exotic injection of energy into the intergalactic medium (IGM).
Late time probes (sensitive at $z\ll 1000$), including the
Lyman-$\alpha$ forest in Quasi-Stellar Objects (QSO) spectra and the
21cm signal of the hyperfine HI transition, are expected to be
particularly efficient in testing late time energy injection. They are
sensitive to the IGM temperature, $T_{\rm k}$, and, consequently, to
exotic energy injections transferred to the IGM in the form of
heating~\cite{Evoli:2014pva,Lopez-Honorez:2016sur}. In particular, the
Lyman-$\alpha$ forest sensitivity to $T_{\rm k}$ at redshifts $z\sim
4-6$ disfavors~\cite{Liu:2020wqz} lifetimes $\tau\sim 10^{-25}$~s for DM
decays into electron-positron pairs for $m_{\rm DM}<$ MeV$/c^2$. On
the other hand, the 21cm signal will be sensitive to $T_{\rm k}$ at
even earlier times, during the so-called Cosmic Dawn (CD) of galaxies
($z\sim$ 10-20) and the epoch of reionisation (EoR, $z\sim$
5-10). Interestingly, the relative dearth of galaxies during the CD
should make it easier to isolate an additional heating contribution
from DM decay or annihilation.

In these proceedings we briefly summarize our recent
work~\cite{Facchinetti:2023slb} on the imprint of DM decays on the
cosmological 21cm signal and the first Fisher matrix forecasts on DM
lifetime constraints for the Hydrogen Epoch of Reionisation Array
(HERA) telescope.  The latter was designed to measure the 21cm power
spectrum at a high signal to noise ratio (S/N).  HERA has completed
deployment \cite{HERA:2021bsv} and is currently analysing data from an
extended observational campaign.  An initial observational result
performed with $71$ antennas (out of the total 331) and only 94 nights
of measurement has already provided the most constraining upper bounds
on the 21 cm power spectrum at redshifts $z=8$ and
10.~\cite{HERA:2022wmy}

Our results, obtained with {\tt exo21cmFAST}~\cite{[exo21cmFAST]} and
{\tt 21cmCAST}~\cite{[21cmCAST]}, clearly indicate that 21cm cosmology
will soon be mature enough to quantitatively probe exotic heating
scenarios like decaying DM models.

\section{21cm signal and DM decay imprint}
\label{sec:signal-DM}

\begin{figure}[t!]
\begin{minipage}{0.99\linewidth}
  \centerline{\includegraphics[angle=0.,width=0.32\linewidth]{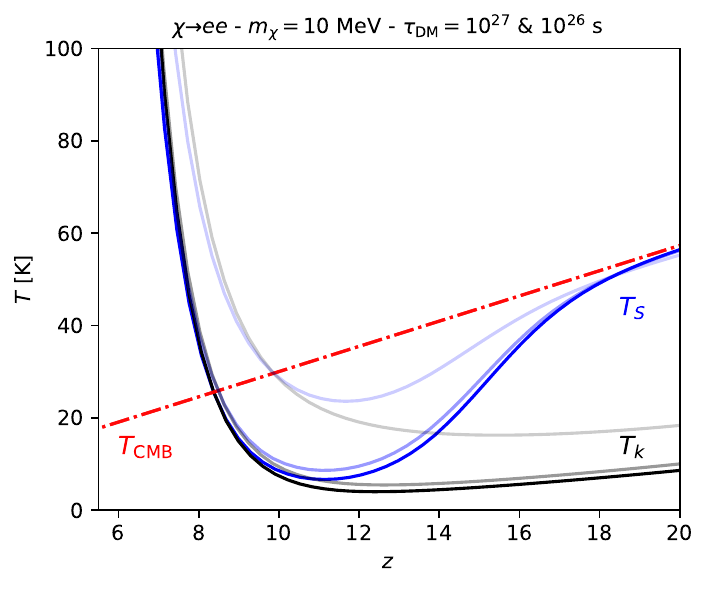}
    \includegraphics[angle=0.,width=0.32\linewidth]{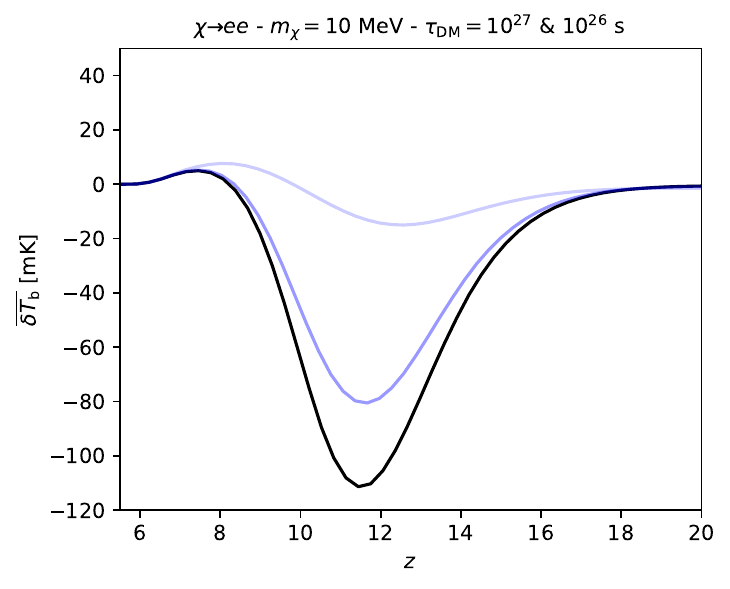}
    \includegraphics[angle=0.,width=0.32\linewidth]{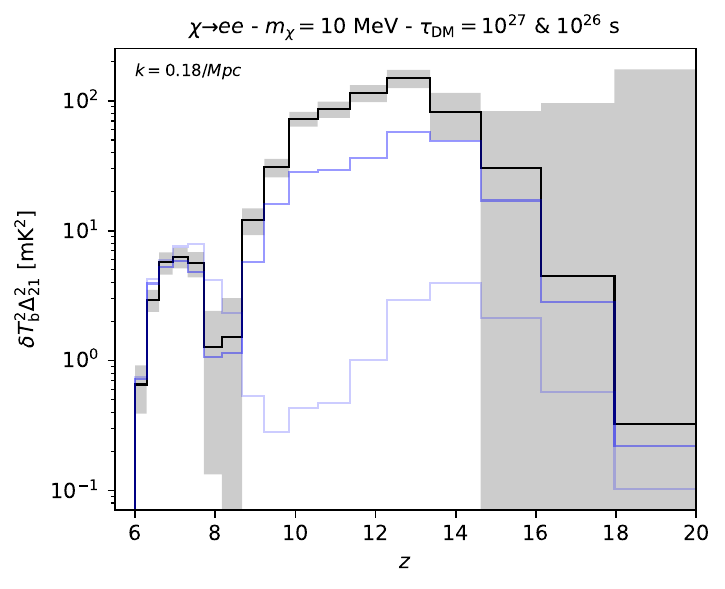}
  }
 \end{minipage}
\caption[]{{\small Imprint of DM decays into a pair of electron
    positrons with 10 MeV$/c^2$ mass and decay rates of $\tau=10^{27}$
    and $10^{26}$s (lightest colored line correspond to longest
    lifetime) compared to a fiducial scenario considering a single
    population of galaxies (darkest lines). In the figures, the
    temperatures (left), the global 21cm signal (central) and the
    power spectrum (PS, right) are shown as a function of the redshift $z$
    in the redshift range of interest. The PS, displayed in $z$ bins
    compatible with HERA specificities, is shown at a scale of
    $k=0.18$/Mpc. The gray area represents the 2 sigma error expected
    from HERA on the fiducial model (black line).  }}
\label{fig:TdTP}
\end{figure}

The redshifted cosmic 21cm signal, arising from the hyperfine
spin-flip transition of neutral hydrogen, can be seen in emission or
in absorption compared to the radio background. Here we fix the latter
to the CMB whose temperature is denoted by $T_\mathrm{CMB}$. The
differential brightness temperature of the 21cm signal, $\delta T_\mathrm{b}$,   shows the
following dependence
\begin{equation}
		\delta T_\mathrm{b} \propto 20\mathrm{mK}
                \left(1-\frac{\mathrm{T_\mathrm{CMB}}}{T_\mathrm{S}}\right)
                x_{\mathrm{HI}}
 \label{eq:deltaTb}
\end{equation}
in the the spin temperature, denoted as $T_{\rm S}$, and in the
neutral fraction, $x_{\rm HI}$.  The spin temperature quantifies the
relative occupancy of the two hyperfine levels of the ground state of
neutral hydrogen. At the redshifts of interest for our study ($z\sim
6-30$) and our assumed astrophysics model, it is obtained from the
equilibrium balance of absorption/emission of 21cm photons from/to the
CMB background, coupling $T_{\rm S}$ to $T_{\rm CMB}$, and resonant
scattering of Lyman-$\alpha$ photons, coupling $T_{\rm S}$ to the gas
temperature $T_{\rm k}$. When the spin temperature is coupled to IGM
gas kinetic temperature through the latter process, $T_{\rm S}$ can
differ from the CMB temperature $T_\mathrm{CMB}$ in
eq.~(\ref{eq:deltaTb}) and the redshifted 21cm signal is seen in
absortion or emission w.r.t. the CMB, see e.g. the left and central
plots of Fig.~\ref{fig:TdTP}. The spatial variation of IGM properties
leads to fluctuations in the 21cm signal. In what follows, we refer to
the 21cm global signal, $\overline{ \delta T_b}$, as the sky averaged
brightness temperature while the 21cm power spectrum (PS),
$\overline{\delta T_b^2} \Delta_{21}^2$, that we will use to derive
our sensitivity curves, is obtained from:
\begin{equation}
    \overline{\delta T_b^2} \Delta_{21}^2(k,z)=\overline{\delta T_b^2(z)} \times \frac{k^3}{2\pi^2} P_{21}(k,z)
    \label{eq:PS}
\end{equation}
where $P_{21}$ is defined as $\langle \tilde\delta_{21} ({\bf k},z)
\tilde\delta_{21} ({\bf k'},z) \rangle= (2\pi)^3 \delta^D({\bf k}-
            {\bf k'}) P_{21}(k,z)$ with $\langle \rangle$ the ensemble
            average, $\bf k$ the comoving wave vector, and
            $\tilde\delta_{21} ({\bf k},z)$ the Fourier transform of
            $\delta_{21} ({\bf x},z)= {\delta T_b}({\bf
              x},z)/\overline{\delta T_b}(z) - 1$. An example of 21cm
            PS is shown in the right panel of Fig.~\ref{fig:TdTP}.

Dark matter energy injection in the IGM, from e.g. annihilations or
decays, induces extra heating, ionization and excitation of the IGM
that leave a footprint in the 21cm signal. In the case of dark matter
decays, the fraction of energy {\it injected} per redshit per unit volume
and baryons, is given by:
\begin{equation}
\frac{1}{n_b}\left( \frac{d E}{dVdz}\right)_{\rm injected}=
   \, \frac{\rho_{\rm DM} c^2}{n_b (1+z) H} \frac{1}{\tau}\,,
   \label{eq:dEdVdtinj}
\end{equation}
where we have assumed that the lifetime satisfy $\tau \gg t_U$, with
$t_U$ the age of the universe, $c$ is the speed of light, $n_b$ the
baryon number density, $\rho_{\rm DM}$ the DM energy density and $H$
the Hubble rate. Below we also use the DM decay rate, $\Gamma\equiv
1/\tau$, to characterize DM energy injection. The fraction of energy
injected of eq.~(\ref{eq:dEdVdtinj}) scales as $(1+z)^{-5/2 }$ in a
matter ($(1+z)^{-3}$ in a radiation) dominated era. From the negative
powers of $(1+z)$ one can expect that DM decays shall leave a stronger
impact on cosmological observables at relatively late times (at $z\ll
1000$) including e.g. 21cm Cosmology probes. In contrast, DM
annihilation through s-wave annihilations is already strongly
constrained by CMB observations~\cite{Planck:2018vyg}. For this reason, we
focus on DM decays instead of DM annihilation for a first forecast of
21cm constraints on DM energy injection in the IGM.

For $z<30$, DM energy injection can dominate the IGM heating at early
times. In Fig.~\ref{fig:TdTP} we illustrate the imprint of DM decays
into a pair of electron positrons with 10 MeV$/c^2$ mass and decay
rates of $\tau=10^{27}$ and $10^{26}$s (lightest colored line
corresponds to longest lifetime) on the gas (blue lines) and spin
temperatures (gray lines) in the left plot, on the global 21cm signal
(blue lines) in the central plot and on the PS (blue lines) in the
right plot. In all cases we consider an astrophysics model
involving a single type of galaxies, refered to as atomic-cooling
galaxies (ACGs), that we have observed at late times.

For the scenarios displayed in Fig.~\ref{fig:TdTP}, the DM heating
dominates for $z>10$--15. From eq.~(\ref{eq:dEdVdtinj}), DM decays
shall give rise to a stronger energy injection in the IGM for shorter
lifetimes (lighest blue colored curves). At fixed mass, a shorter
lifetime is then expected to induce a stronger heating or a larger
increase of $T_k$ and $T_S$ at $z>10$--15 for the scenario illustated
in in Fig.~\ref{fig:TdTP}. This also implies a shallower absorption
trough in the global signal. Both effects are well visible in the left
and central panels of Fig.~\ref{fig:TdTP}. DM decays also induce a
more uniformly heated IGM at early times, which can decrease the
large-scale 21cm power during Cosmic Dawn compared to galaxy-only
heating~\cite{Evoli:2014pva,Lopez-Honorez:2016sur} as illustrated in
the right panel of Fig.~\ref{fig:TdTP}.  Let us mention that the
impact of DM decay on the late-time 21cm power ($z<9$ for our scenario
displayed in Fig.~\ref{fig:TdTP}) is far more modest compared to
astrophysics sources of X-rays.~\cite{Facchinetti:2023slb}  Given that
experiments such as HERA will be able to probe a large range of
redshifts and scales, we expect them to be able to disentangle these
two sources of heating. From the HERA sensitivity estimate shown in
Fig.~\ref{fig:TdTP} with a light gray area, we can expect that HERA
shall be able to probe lifetimes up to $~10^{27}-10^{28}$ s,
surpassing the sensitivity from CMB and Lyman-$\alpha$ probes.

\section{Results}
\label{sec:results}

We have performed the first Fisher matrix
analysis~\cite{Facchinetti:2023slb} to evaluate the best lower limits
that are expected to be set on the dark matter lifetime $\tau$ by the
HERA experiment within two possible astrophysics scenarios and fixed
dark matter mass $m_\chi$.  The first scenario considers only
radiation from \popII{}-dominated ACGs. ACGs have been observed and
their stellar to halo mass relation is well constrained by UV
luminosity functions. On the other hand, we have also considered the
case of a more complete, yet more complex, astrophysics model where we
consider both ACGs and molecular-cooling galaxies (MGCs). The latter
scenario involves more sources of uncertainties as MGCs properties are
yet to be determined.  MGCs would correspond to the very first
galaxies, hosted by so-called minihalos (with a virial mass $< 10^8
M_\odot$) that are expected to predominately host Population-III
stars.   In particular, MGCs can heat the medium earlier than AGCs and
the DM imprint becomes less easy to untangle from astrophysics. This
is expected to mitigate the constraints on exotic sources of
heating.

The Fisher matrix analysis including AGCs only (AGCs+MGCs)
involves a set of 8 (11) astrophysical parameters and one dark matter
parameter which is the decay rate, $\Gamma$. In the case of AGC only
(AGC\& MGCs), it appears that one of the most degenerate parameter
with $\Gamma$ is the integrated soft-band luminosity per star
formation rate from AGCs (from MGCs), denoted with $L_X^{II}$
($L_X^{III}$). The latter essentially determines the amplitude of X
ray heating from cosmic dawn galaxies.

\begin{figure}[t!]
\begin{minipage}{0.99\linewidth}
\centerline{\includegraphics[angle=0.,width=0.99\linewidth]{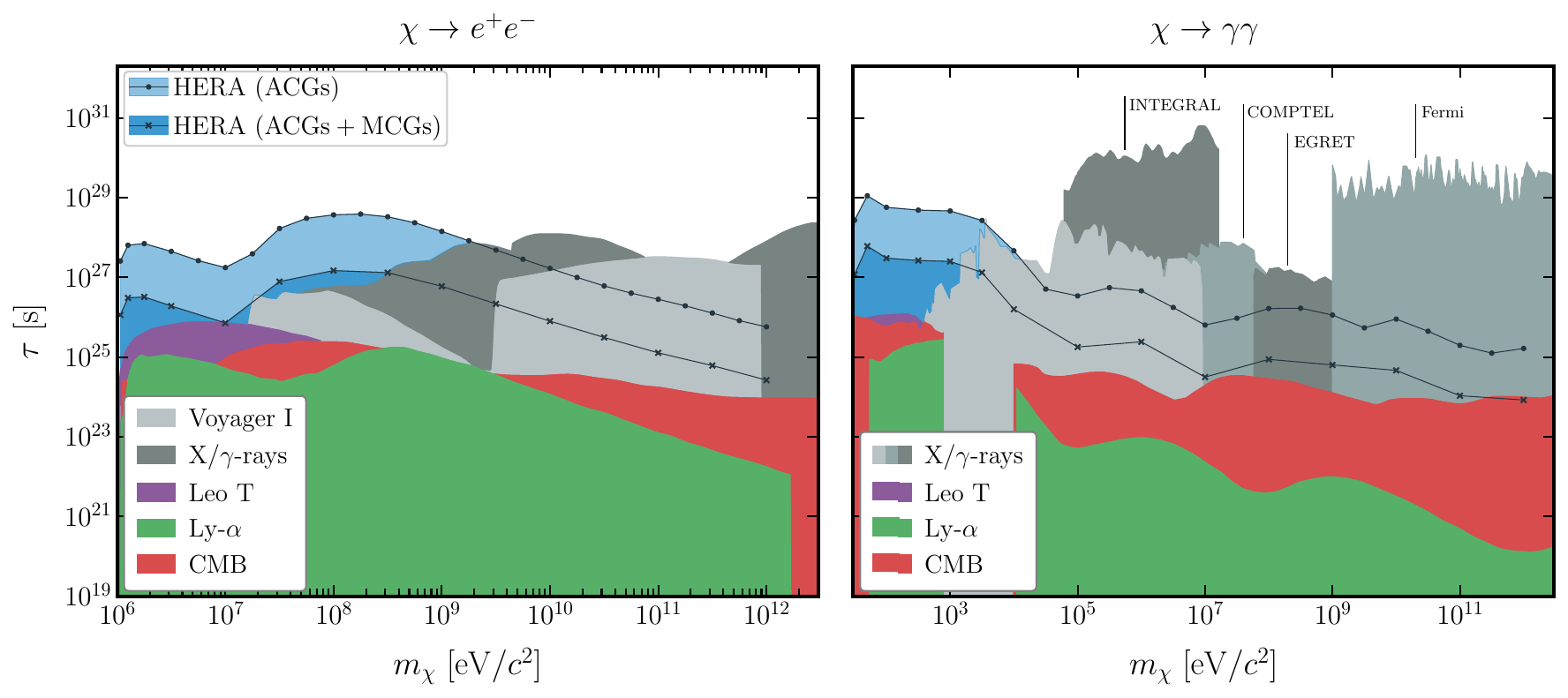}}
\end{minipage}
\caption[]{{\small Updated constraints on the dark matter lifetime (at 95\% level) for decay into an electron/positron pair (left panel) and photons (right panel). We superpose the forecasts for the HERA telescope assuming \popII{}-dominated ACGs only (solid dark line with round markers), or \popII{}-dominated ACGs  + \popIII{}-dominated MCGs with $\log_{10}(L_X^{\rm II,III}) = 40$  (solid dark line with crosses) with existing cosmological constraints. }}
\label{fig:cons}
\end{figure}

Our main results, presented in Fig.~\ref{fig:cons}, display the lower
bound at a 95\% confidence level (CL) on the lifetime of dark matter
derived from our Fisher matrix forecasts based on HERA specifications.
Black lines with bullets are obtained for one single population of
galaxies (AGCs) while black lines with crosses assume ACGs+MCGs. The blue
area below these curves are excluded at 95\% CL. When considering the
ACGs+MCGs scenario, heating from galaxies competes with DM heating
earlier and the DM heating parameters become more difficult to
constrain.  The lower bound on the DM life time becomes thus less
stringent in ACGs+MCGs scenario (crosses) than in the AGCs only case
(bullets). For decays into $e^+e^-$, a few 100 MeV$/c^2$ DM gets the
most stringent 21cm bounds on $\Gamma$ while, for decays into photons,
it is the case for the lowest DM masses (with $m_\chi<$MeV$/c^2$).

  In Fig.~\ref{fig:cons}, we also show the lower bounds arising from
  cosmology and astrophysics probes such as Lyman-$\alpha$
  forest~(green), CMB~(red) and Leo T~(purple).  21-cm measurements
  with HERA could improve by up to 3 orders of magnitude the current
  limits on the DM lifetime, when considering \popII{}-dominated ACGs
  only (black line with bullets). In addition, we also show the
  existing bounds from indirect dark matter searches including
  constraints from the Voyager I observation of cosmic rays, and from
  X- or $\gamma$-ray experiments such as INTEGRAL/SPI, COMPTEL, EGRET
  and Fermi.\footnote{More competitive constraints from XMM Newton
  have recently been published~\cite{DelaTorreLuque:2023olp}, see talk
  of P. De la Torre Luque in Moriond VHEPU talks.} We have found that
  existing constraints for dark matter heavier than 1 GeV$/c^2$ (or
  100keV$/c^2$) for decays into $e^+e^-$ (or photons) remain
  competitive and our 21cm forecast for 1000 hours of HERA observation
  is unlikely to improve these limits in the higher DM mass range.

\begin{figure}[t!]
\begin{minipage}{0.99\linewidth}
\centerline{\includegraphics[angle=0.,width=0.55\linewidth]{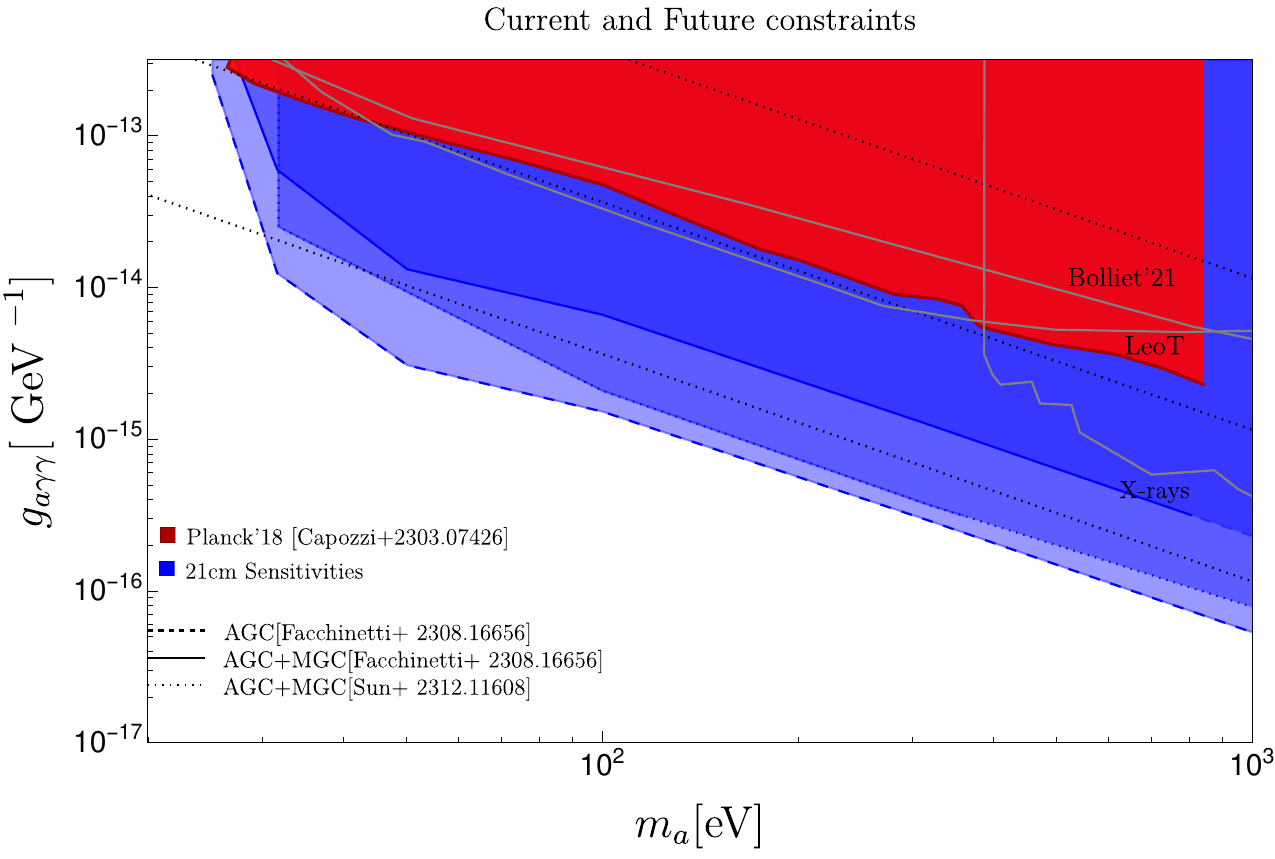}}
\end{minipage}
\caption[]{{\small Present and future bounds on the ALP
    $(m_a,g_{a\gamma\gamma})$ parameter space. These include existing
    95\%~CL cosmology constraints from Planck 2018 assuming a $\tanh$
    reionization~\cite{Capozzi:2023xie} in red and expected future
    HERA sensitivity estimate in blue from our
    work~\cite{Facchinetti:2023slb} and a recent analysis accounting
    for inhomogeneous injection~\cite{Sun:2023acy}}. The dotted black
  lines indicate, from top to bottom, axion life times of $\tau
  =10^{24}, 10 ^{26 }$ and $10^{28}$ s.}
\label{fig:consALP}
\end{figure}

In Fig.~\ref{fig:consALP}, we project the HERA sensitivity for
$\chi\to \gamma \gamma$ in the plane $(m_a,g_{a\gamma\gamma})$
corresponding to the mass and coupling to photons of DM, in the form
of an axion like particle (ALP) decaying to two photons. For the
latter purpose we have assumed that the axion decay rate scales as
$\Gamma_a=\frac{g_{a\gamma\gamma}^2}{64\pi} m_a^3 $. The red region is
excluded at 95\%~CL by Planck 2018 data assuming a $\tanh$
reionization model~\cite{Capozzi:2023xie} while the blue region
illustrates the 95\% CL sensitivity of HERA. The later can probe
$g_{a\gamma\gamma}$ couplings smaller by more than order of magnitude
than Planck data and can also improve on X-ray searches in the $\sim$
MeV masses (all references for other (astro-)particle bounds in gray
colors can be found in~\cite{Capozzi:2023xie}). Notice that our
results, shown with the continuous and dashed blue contours, were
obtained considering an homogeneous DM energy injection. The dashed
(continuous) contour corresponds to the ACGs only (AGCs+MGCs) limits
represented with solid dark line with round markers (crosses) in the
left panel of Fig.~\ref{fig:cons}. Our results can be compared to the
dotted blue contours from a more recent analysis accounting for
inhomogeneous injection~\cite{Sun:2023acy} and a yet different
astrophysics background involving AGCs+MGCs. It appeared that, when
accounting for inhomogeneities, the sensitivity reach of HERA only
marginally weakens compared to the homogeneous treatment even though
the detailed 21cm fluctuation maps might differ more significantly
depending on the decaying DM lifetimes.\footnote{ This conclusion
could be expected as the differential brightness contrast sourced by
inhomogeneous DM energy injection is expected to be enhanced but the
potential increase in sensitivity is counterbalanced by increasing
degeneracies with astrophysics parameters.~\cite{Sun:2023acy}}  To
conclude, Fig.~\ref{fig:consALP} emphasizes again the improvement in
probing decaying DM candidates that can be expected from 21cm
cosmology experiments probing the 21cm power spectrum.

\section{Conclusion}

As is well known, DM decays give rise to a relatively late time ($z\ll
1000$) energy deposition into the medium. This makes late time probes,
such as Lyman-$\alpha$ forest or 21cm cosmology, very interesting
targets to detect the DM imprint.  In this work, we focus on the
effect of DM on the 21cm signal power spectrum and prospects for
constraints on the DM lifetime by the HERA interferometer. This
telescope will enable us to explore a vast range of redshifts,
stretching from the Epoch of Reionization to Cosmic Dawn, with
exceptional precision. This capability is of paramount importance
because DM is not the sole contributor to the heating process, and it
is crucial to distinguish its distinct signature from that generated
by X-rays emitted from the first galaxies.

Our Fisher matrix forecasts for 21cm power spectrum measurements
sensitivity to DM decays are very promising. Our results are summarized
in Fig.~\ref{fig:cons} and projected in the case of ALPs in the
relevant mass range in Fig.~\ref{fig:consALP} for these
proceedings. When considering the minimal astrophysics scenario (AGC
only), HERA is expected to improve on existing cosmology constraints
(from CMB and Lyman-$\alpha$ probes) on the DM lifetime by up to 3
orders of magnitude.  This corresponds to more than one order of
magnitude improvement on the ALP photon coupling $g_{a\gamma\gamma}$ in
the mass range of $m_a\sim 10$ keV to MeV. We also compare these
prospects to the case where the astrophysics model includes both
\popII{}-dominated ACGs and \popIII{}-dominated MCGs. Similarly to DM,
MCGs is expected to give rise to a new source of IGM heating before
\popII{}-dominated ACGs light on, partially drowning the DM
signal. Nevertheless, even in the latter case, HERA can improve on
existing cosmology constraints by a factor of 10 to 100. Finally,
compared to existing $\gamma$-ray and cosmic-ray limits, HERA is
expected to be a key player in constraining DM candidates decaying to
$e^+e^-$ in the mass range $m_\chi <$ 2 GeV$/c^2$. For decays into
photons, HERA improves on other searches in the low mass range for
$m_\chi<$ few MeV$/c^2$.

\section*{Acknowledgments}
These proceedings mainly summarize the work~\cite{Facchinetti:2023slb} done
in collaboration with G. Facchinetti, Y. Qin and A. Mesinger where
all missing details and references (due to restricted number of pages) can be
found. LLH thank the organizers of Moriond Electroweak Interactions \&
Unified Theories 2024 and of the Corfu Summer Institute 2023 for the
the invitation and the very nice conferences.  She is supported by the
Fonds de la Recherche Scientifique F.R.S.-FNRS through a research
associate position and acknowledges support of the FNRS research grant
number F.4520.19, the ARC program of the Federation Wallonie-Bruxelles
and the IISN convention No. 4.4503.15.


\bibliographystyle{unsrt}    


\end{document}